\documentclass[9pt,sigconf]{acmart}
\renewcommand\footnotetextcopyrightpermission[1]{}

\usepackage{textcomp}
\usepackage[table]{xcolor}

\usepackage{comment}
\usepackage{booktabs}
\usepackage{multirow}
\usepackage{array}
\usepackage{svg}
\usepackage[labelformat=simple]{subcaption}

\usepackage{amsmath}   % 数学符号支持

\usepackage[mathcal]{eucal}
\usepackage[ruled,vlined,noend]{algorithm2e}
\usepackage[]{algpseudocode}                               
\usepackage{enumitem}

\usepackage{tabularx}

\usepackage{multirow}
\usepackage{xcntperchap}
\usepackage{tikz}

\usepackage{fancyhdr} 
\AtBeginDocument{%
  }

\setcopyright{acmlicensed}
\copyrightyear{2018}
\acmYear{2018}
\acmDOI{XXXXXXX.XXXXXXX}
\acmConference[Conference acronym 'XX]{Make sure to enter the correct
  conference title from your rights confirmation email}{June 03--05,
  2018}{Woodstock, NY}
\acmISBN{978-1-4503-XXXX-X/2018/06}

\begin{document}

\title{On the Vulnerability of\\ FHE Computation to Silent Data Corruption}
% \author{Anonymous authors}
 % \affiliation{
 %   \institution{Paper under double-blind review}
 % }
\renewcommand{\shortauthors}{Anonymous Author, et al.}

\author{Jianan Mu}
\authornote{Both authors contributed equally to this research.}
\email{mujianan@ict.ac.cn}
\affiliation{%
  \institution{Institute of Computing Technology}
  \institution{Chinese Academy of Sciences}
  \city{Beijing}
  \country{China}
}
\author{Ge Yu}
\authornotemark[1]
\email{yuge23s@ict.ac.cn}
\affiliation{%
  \institution{School of Advanced Interdisciplinary Sciences}
  \institution{University of Chinese Academy of Sciences}
  \city{Beijing}
  \country{China}
}

\author{Zhaoxuan Kan}
\email{kanzhaoxuan23z@ict.ac.cn}
\affiliation{%
  \institution{Institute of Computing Technology}
  \institution{Chinese Academy of Sciences}
  \city{Beijing}
  \country{China}
}

\author{Song Bian}
\email{sbian@buaa.edu.cn}
\affiliation{%
  \institution{School of Cyber Science and Technology}
  \institution{Beihang University}
  \city{Beijing}
  \country{China}
}

\author{Liang Kong}
\email{kongrenky@163.com}
\affiliation{%
  %\institution{Ant Group}
  \city{Beijing}
  \country{China}
}

\author{Zizhen Liu}
\email{liuzizhen@ict.ac.cn}
\affiliation{%
  \institution{Institute of Computing Technology}
  \institution{Chinese Academy of Sciences}
  \city{Beijing}
  \country{China}
}

\author{Cheng Liu}
\email{liucheng@ict.ac.cn}
\affiliation{%
  \institution{Institute of Computing Technology}
  \institution{Chinese Academy of Sciences}
  \city{Beijing}
  \country{China}
}

\author{Jing Ye}
\email{yejing@ict.ac.cn}
\affiliation{%
  \institution{Institute of Computing Technology}
  \institution{Chinese Academy of Sciences}
  \city{Beijing}
  \country{China}
}

\author{Huawei Li}
\email{lihuawei@ict.ac.cn}
\affiliation{%
  \institution{Institute of Computing Technology}
  \institution{Chinese Academy of Sciences}
  \city{Beijing}
  \country{China}
}

\begin{abstract}
Fully Homomorphic Encryption (FHE) is rapidly emerging as a promising foundation for privacy-preserving cloud services, enabling computation directly on encrypted data. 
As FHE implementations mature and begin moving toward practical deployment in domains such 
as secure finance, biomedical analytics, and privacy-preserving AI, a critical question 
remains insufficiently explored: how reliable is FHE computation on real hardware?
This question is especially important because, compared with plaintext computation, FHE incurs much higher computational overhead, making it more susceptible to transient hardware faults. Moreover, data corruptions are likely to remain silent: the FHE service has no access to the underlying plaintext, causing unawareness even though the corresponding decrypted result has already been corrupted.
To this end, we conduct a comprehensive evaluation of SDCs in FHE ciphertext computation.
Through large-scale fault-injection experiments, we characterize the vulnerability of FHE to transient faults, and through a theoretical analysis of error-propagation behaviors, we gain deeper algorithmic insight into the mechanisms underlying this vulnerability.
We further assess the effectiveness of different fault-tolerance mechanisms for mitigating these faults.

\end{abstract}
\maketitle

\section{Introduction}

% 图1展示了FHE重要且极具吸引力的应用场景：privacy-preserving 云服务：在FHE密文计算技术的使能下，allows clients to outsource computations to untrusted servers without exposing sensitive information.

\begin{figure}[t]
    \centering
    %\includegraphics[width=0.8\linewidth]{fig/fig_1_1117.png}
    %\includesvg[inkscapelatex=false,width=1.0\linewidth]{fig/fig1.svg}
    \includegraphics[width=1.0\linewidth]{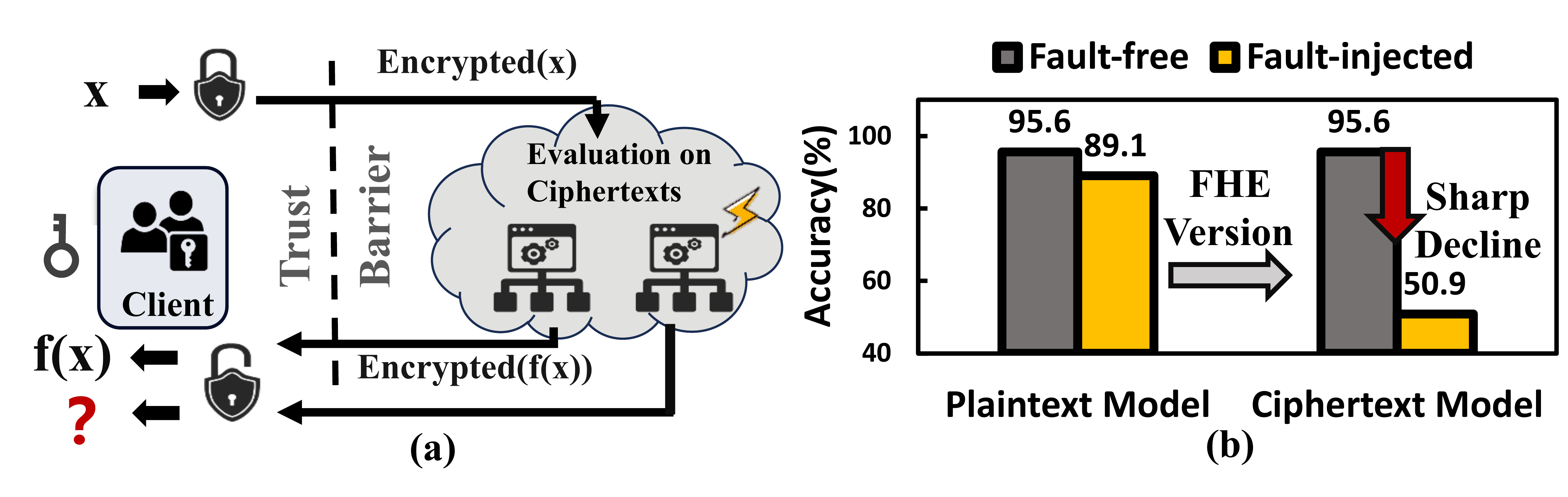}
    \caption{ (a) Typical application scenarios of FHE. (b) An FHE-based cancer detection service showing that accuracy degrades sharply under a single-bit hardware fault.}
    \label{fig:fhe_app}
    \vspace{-0cm}
\end{figure}

Fully Homomorphic Encryption (FHE)~\cite{BFV,TFHE,CKKS} is an emerging cryptographic technique that enables homomorphic computations directly on encrypted data.  
As shown in Fig.~\ref{fig:fhe_app}(a), it allows clients to outsource computations to untrusted servers without exposing sensitive information, since all data remains encrypted throughout storage, transmission, and processing.  
This ideal privacy-preserving property has attracted substantial investment from both industry and government, driving FHE’s evolution from a theoretical construct to an emerging foundation for security-critical services such as encrypted finance, bioinformatics, and AI inference~\cite{FHE_service,FHE_ai_service, CKKSCNN,HELR,LoLa}.  
Among existing FHE schemes, the CKKS~\cite{CKKS} has received particular attention due to its support for vectorized SIMD-style computation and approximate arithmetic over complex numbers, making it especially well suited for AI and numerical applications.
While most existing efforts toward practical FHE deployment concentrate on improving the performance of ciphertext computation, the \textbf{reliability of FHE operations on real-world hardware}—their behavior under transient faults—remains a critical yet largely overlooked dimension.

% While recent research efforts have mainly focused on improving the performance of ciphertext computation, the \textbf{reliability of FHE operations on real-world hardware}—that is, their behavior under transient faults remains a critical yet largely unexplored issue.

Real-world hardware inevitably experiences transient faults that may cause \textit{silent data corruption (SDC)}—erroneous results without visible system failures~\cite{papadimitriou2023silent}.
Under such random-fault threats, ciphertext-based cloud services face greater reliability challenges than plaintext computations due to the unique nature of encrypted execution.
\textbf{First, hardware faults are more likely to occur:} because FHE evaluation is orders of magnitude more expensive than plaintext processing (often exceeding a $10^{4}\times$ slowdown), the substantially longer execution time inherently increases the probability of transient faults during computation.
\textbf{Second, data corruption is more likely to remain silent:} while plaintext systems can often flag abnormal values (e.g., \texttt{NaN}, out-of-range numbers, or unexpected signals), an FHE server has no access to the underlying plaintext, and ciphertexts appear pseudorandom. Consequently, the server cannot determine whether a returned ciphertext is valid or corrupted.
\textbf{Finally, SDCs may fundamentally undermine trust in ciphertext-based services:} FHE is typically deployed in high-value and privacy-critical domains such as secure finance and biomedical analytics. Returning an SDC-corrupted ciphertext to the client while the server is unaware of the corruption may therefore lead to severe consequences.

This raises an important issue: \textbf{how do hardware faults during ciphertext computation affect the final decrypted result?}
At first glance, FHE may seem capable of absorbing certain perturbations.
Modern lattice-based FHE schemes are built upon the Learning With Errors (LWE) or Ring-LWE assumption, under which each ciphertext includes a small, mathematically structured noise term.
This algorithmic noise is carefully bounded, and decryption succeeds as long as the accumulated noise remains within its prescribed limit.
However, such tolerance applies only to the \textit{specific form, distribution, and magnitude of noise} restricted by the cryptographic construction.
In contrast, hardware-introduced noise becomes uncontrolled in both distribution and magnitude through the complex propagation process of ciphertext computation.
This gap motivates a  finer-grained examination of how hardware-induced errors propagate through ciphertext computation and ultimately affect the decrypted result.
To examine this issue, we conduct a single-bit transient fault injection experiment on a CKKS-based encrypted neural network performing binary cancer classification using the Breast Cancer Wisconsin~\cite{breast_cancer_wisconsin} dataset (Fig.~\ref{fig:fhe_app}(b)).
%In Fig.~\ref{fig:fhe_app}(b), 
We compare the accuracy degradation of plaintext inference under injected faults with that of ciphertext inference after decryption.
The results show a clear divergence in behavior: while plaintext inference retains a certain level of robustness to single-bit faults, the accuracy of ciphertext inference exhibits a sharp decline.
This comparison shows that FHE computations are inherently intolerant to transient faults, underscoring the urgent need for a systematic evaluation of hardware faults in ciphertext computation.

In this paper, we provide a systematic evaluation of SDCs in FHE computation. First, we perform large-scale fault-injection experiments on CKKS ciphertext operations and observe that even a single-bit transient fault during encrypted computation can lead to erroneous decrypted outputs, while remaining 
undetectable to the server.
Next, we theoretically analyze how errors propagate and amplify across slot and bit levels throughout the CKKS computation, revealing the structural mechanisms that make FHE computation highly vulnerable.
Finally, we evaluate the performance of redundant
and checksum-based fault-tolerance methods for this problem.
Our main contributions are summarized as follows:
\begin{itemize}
  \item \textbf{Systematic study.} 
We conduct a systematic investigation of an interesting and important problem: hardware-induced SDCs in CKKS computation, combining large-scale fault-injection experiments with a structural analysis of the underlying algorithms.
  \item \textbf{Large-scale fault-injection experiments.} 
We inject a large number of random transient faults to characterize the vulnerability and error magnitude of FHE ciphertext computation. 
\item \textbf{Fault-tolerance evaluation.}
 We evaluate redundant and
checksum-based fault-tolerance algorithms for the FHE pipeline
and quantify their effectiveness and overhead.
\end{itemize}

% 本文的主要创新性在于针对FHE计算中SDC的脆弱性的全面分析,并通过算法结构分析揭示其脆弱性的底层机制.基础容错算法实现和评估是为了保证研究完整性,而不是本文的主要目标.

% 1. 系统性研究
% 针对CKKS计算中硬件引入的SDC,我们进行了一个包括大量实验和算法结构分析的系统性研究.

% 2. 故障注入分析
% 我们注入了大量随机故障,刻画了FHE计算对此的脆弱性和误差范围.

% 3. error传播的理论分析
% 我们基于算法结构特性,进行了error传播的细粒度分析,展示了error在FHE计算中的slot和bit level的放大效果,提供了对于FHE计算对硬件引入SDC脆弱性的更深入理解.

% In this paper, we propose a systematic evaluation of SDCs in FHE computation.
% 首先, we conduct fault-injection experiments on CKKS ciphertext computations and observe that even a single-bit transient fault during computation on ciphertext can frequently cause a result leads to erroneous decryption outputs, while the server is unaware.
% 进一步的, we theoretically analyze how faults propagate and amplify across slot- and bit-levels through the FHE computation flow, providing a deeper understanding of the underlying failure mechanisms.
% 为了完整性,我们还评估并报告了baseline容错算法的效果和开销.
% 最后，我们评估了不同的容错算法，包括基于冗余，和基于check sum（我们整合不同多项式check sum并扩展到FHE计算流）对于FHE密文计算故障的表现。

\section{Preliminary}

% \begin{figure}[t]
%     \centering
%     \includegraphics[width=0.4\linewidth]{fig/muli-level_data_operator.png}
%     \caption{The multi-level structure of data and operators in FHE.}
%     \label{fig:fhe_app}
% \end{figure}

\subsection{Silent Data Corruptions}

Silent Data Corruptions (SDCs) are hardware-induced failures where systems produce incorrect results without raising logs, exceptions, or error reports. They may arise from transient faults (often manifesting as soft errors), subtle manufacturing defects, or escaped design bugs, causing processors to miscompute operations or propagate incorrect values. Because current error-reporting mechanisms fail to capture such behaviors, SDCs often remain invisible until they silently spread across applications or entire datacenter fleets~\cite{papadimitriou2023silent}. 

At the hardware level, protection typically relies on guardbands, error-correcting codes (ECC), and redundant execution~\cite{asgari2023structural, udipi2012lot}. Guardbands and ECC provide effective safeguards for memory, while redundant execution can improve logic reliability, but these techniques are insufficient to eliminate silent errors and often introduce significant overhead. Recent reports from Meta, Google, and Alibaba confirm that CPUs and AI accelerators continue to experience SDCs in production despite extensive validation~\cite{ma2024proactive,wang2023understanding}. To complement hardware safeguards, algorithm-based fault tolerance (ABFT) introduces redundancy at the algorithmic level~\cite{zhai2023ft}. For example, checksum encoding in matrix operations enables efficient detection and correction of corruptions, offering resilience without the prohibitive costs of hardware-only approaches.

\subsection{CKKS Scheme} 
\label{sec:CKKS-prelim}

CKKS is a representative arithmetic homomorphic encryption scheme that supports floating-point computation and efficient vectorized SIMD operations.
Its ciphertext evaluation is carried out through a data-oblivious, control-flow–independent sequence of polynomial-domain operations.
In this work, we focus our fault-injection analysis on CKKS computations.
In the following, we introduce the CKKS encryption and decryption procedures, its ciphertext data structure, and the operators involved in ciphertext computation.

\subsubsection{Encryption and Decryption}

In the CKKS scheme, a user message $m$ is represented as a complex-valued vector.  
Before encryption, the message $m$ is first encoded into a plaintext polynomial 
$\mathbf{pt} \in \mathcal{R}_Q$, 
where $\mathcal{R}_Q = \mathbb{Z}_Q[X]/(X^N + 1)$ denotes the polynomial ring under modulus $Q$.  
The encryption process transforms the plaintext $\mathbf{pt}$ into a ciphertext 
$\mathbf{ct} = (c_0, c_1) \in \mathcal{R}_Q^2$ as $\mathbf{ct} = (c_0, c_1) = (-a \cdot s + \mathbf{pt} + e,\, a) \ mod \ Q,$ where $s$ is the secret key polynomial, $a$ is a uniformly sampled random polynomial from $\mathcal{R}_Q$, and $e$ is chosen from some distribution $\chi_{noise}$ that provides semantic security based on the Ring Learning With Errors (RLWE) assumption~\cite{LWE_error}.
% is a small Gaussian error polynomial 

During decryption, the ciphertext is recombined with the secret key as $ c_0 + c_1 \cdot s = \mathbf{pt} + e \pmod{Q}$. After decryption, the intermediate plaintext before decoding is: $\mathbf{pt}' = \Delta \cdot Encode(m) + e$,  where $\Delta$ is the scaling factor that separates high-significance message bits from low-order noise components. During the decoding step, the system divides by $\Delta$ to recover the approximate message: $
\hat{m} =  m + \frac{e}{\Delta}.$
% This division by $\Delta$ effectively truncates the low-bit noise term $\tfrac{e}{\Delta}$, preserving only the high-significance portion of the message while filtering out minor perturbations introduced by encryption or homomorphic computation.  
% Therefore, the scaling factor $\Delta$ not only defines the arithmetic precision of CKKS but also serves as a natural noise-suppression mechanism during decryption.
This division by $\Delta$ effectively truncates the low-bit noise term $\tfrac{e}{\Delta}$, preserving only the high-significance portion of the message while filtering out minor perturbations introduced during the encryption process.
Therefore, the scaling factor $\Delta$ not only defines the arithmetic precision of CKKS but also serves as a natural noise-suppression mechanism during decryption.

\subsubsection{Multi-Level Data Structures}

\begin{figure}[htbp]
    \centering
    %\includegraphics[width=0.4\linewidth]{fig/muli-level_data_operator-1115.png}
    %\includesvg[inkscapelatex=false,width=0.4\linewidth]{fig/fig2.svg}
    \includegraphics[width=0.4\linewidth]{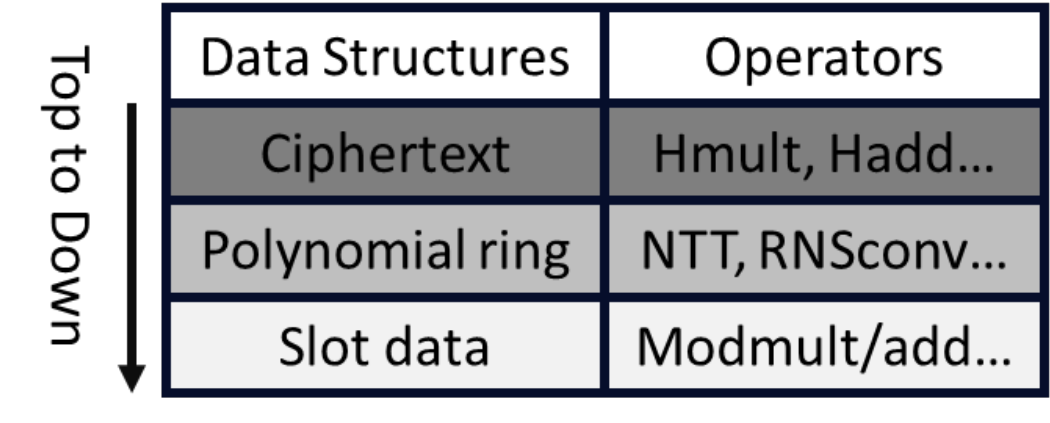}
    \vspace{-0.3cm}
    \caption{Multi-level structure of data and operators in CKKS.}
    \label{fig:fhe_data_op}
    \vspace{-0.3cm}
\end{figure}

As shown in Fig.~\ref{fig:fhe_data_op}, CKKS adopts a hierarchical data organization with three levels: \textbf{ciphertext}, \textbf{polynomial ring}, and \textbf{slot data}.
At the top level, a ciphertext $(c_0, c_1)\in \mathcal{R}_Q^2$ encapsulates the encoded message and noise. Each polynomial lies in $\mathcal{R}_Q=\mathbb{Z}_Q[X]/(X^N+1)$, where the large modulus $Q$ is decomposed by the Residue Number System (RNS) into smaller primes $q_i$ for parallel processing.
% At the bottom level, it is each slot data of the polynomial.
At the bottom level, it is the data in each slot of the polynomial.

\begin{figure*}[t]
    \centering
    %\includegraphics[width=0.8\linewidth]{fig/fig_3_1117.png}
    %\includesvg[inkscapelatex=false,width=0.9\linewidth]{fig/fig3.svg}
    \includegraphics[width=0.9\linewidth]{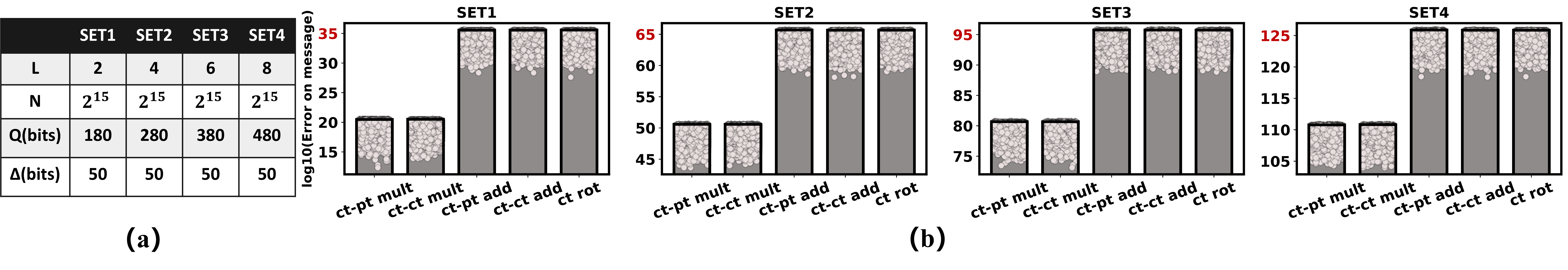}
    \vspace{-0.4cm}
    \caption{ Vulnerability evaluation of 5 CKKS homomorphic operations (ct-pt~mult, ct-ct~mult, ct-pt~add, ct-ct~add, ct~rot): (a) Parameter settings. (b) Slot-level error variation of different parameter settings.}
    \vspace{-0.4cm}
    \label{fig:fhe_operator}
\end{figure*}

\subsubsection{Multi-Level Operators}

% In CKKS, ciphertext computation is composed of hierarchical primitives across different abstraction levels.
% At the \textbf{ciphertext level}, basic operations include ciphertext addition (\textbf{CAdd}), subtraction (\textbf{CSub}), multiplication (\textbf{CMul}), and rotation (\textbf{CRot}). \textbf{CAdd} and \textbf{CSub} are implemented as modular additions and subtractions over polynomials in $\mathcal{R}_Q^2$, while \textbf{CMul} performs polynomial multiplications between underlying plaintexts, producing an intermediate triplet $(d_0,d_1,d_2)$ before relinearization. \textbf{CRot} applies an automorphism transformation $\sigma_r: i \mapsto i \cdot 5^r \bmod N$ to rotate encoded slots. Both \textbf{CMul} and \textbf{CRot} require a subsequent \textit{keyswitch} to map the ciphertext back under the original secret key $s$.
In CKKS, ciphertext computation consists of a hierarchy of primitives across different abstraction levels. At the \textbf{ciphertext level}, basic operations include ciphertext–plaintext addition (\textbf{ct-pt add}), ciphertext–ciphertext addition (\textbf{ct-ct add}), ciphertext–plaintext multiplication (\textbf{ct-pt mult}), ciphertext–ciphertext multiplication (\textbf{ct-ct mult}), and ciphertext rotation (\textbf{ct rot}). Both \textbf{ct-pt add} and \textbf{ct-ct add} correspond to modular additions over polynomials. The \textbf{ct-pt mult} primitive multiplies each ciphertext component by the encoded plaintext polynomial, while \textbf{ct-ct mult} performs a full polynomial multiplication between two ciphertexts, producing an intermediate triplet $(d_0,d_1,d_2)$ before relinearization. \textbf{ct rot} applies an automorphism transformation $\sigma_r: i \mapsto i \cdot 5^r \bmod N$ to rotate encoded slots. Both \textbf{ct-ct mult} and \textbf{ct rot} require a subsequent \texttt{keyswitch} to map the ciphertext back under the original secret key $s$.

At the \textbf{polynomial level}, it invokes three core arithmetic operators: \textbf{Number Theoretic Transform (NTT/INTT)}, \textbf{Basis Conversion (BConv)}, and \textbf{Point Multiplication}.
During decryption, it further performs \textbf{DCRT Interpolation}, which reconstructs a full-modulus polynomial by aggregating the per-modulus residues distributed across the CRT moduli.
% NTT accelerates polynomial multiplication by converting coefficient representations into the evaluation domain through butterfly operations; BConv switches between different RNS bases $(Q \rightarrow P)$ to manage modulus reduction; and Point Multiplication performs element-wise modular operations on decomposed limbs.
At the \textbf{slot level}, computations eventually reduce to modular multiplications and additions on encoded plaintext slots, completing the full homomorphic arithmetic hierarchy from slot to ciphertext.

% 放在导言区：需要 booktabs
% \usepackage{booktabs}

\subsection{Related Works}
\label{related works}

For lattice-based cryptography, existing fault analysis studies have primarily focused on \textit{fault injection attacks} targeting the decryption stage to extract secret keys~\cite{PQC_FaultAttack2023}.  
A very recent arXiv report investigated the reliability of FHE ciphertexts under memory faults~\cite{FHE_MemFault2025}, showing that ECC can protect against single-bit flips in memory. However, it does not examine faults originating from other sources and lacks an in-depth analysis of fault-propagation paths within the FHE computation flow.
Moreover, although there is some client-side verification mechanism after decryption~\cite{OpenFHE}, it remains difficult to provide low-overhead reliability protection for the server without introducing information leakage.

\subsection{Notations}
\label{notations}

The notations used in this paper are listed in Table~\ref{tab:notation}.

\begin{table}[h]
\centering
\scriptsize
\renewcommand{\arraystretch}{1.05}
\begin{tabular}{p{0.20\columnwidth} p{0.72\columnwidth}}
\toprule
\textbf{Symbol} & \textbf{Description} \\ \midrule

$m$, $\mathbf{pt}$, $\mathbf{ct}$ &
message; encoded plaintext; ciphertext $(c_0,c_1)$. \\[2pt]

$\mathbb{Z}_Q$, $\mathcal{R}_Q$ &
Integer ring $\mathbb{Z}_Q$; polynomial ring $\mathcal{R}_Q=\mathbb{Z}_Q[X]/(X^N+1)$. \\[2pt]

$N$, $Q$, $\Delta$, $L$ &
Polynomial degree; ciphertext modulus; CKKS scaling factor; ciphertext multiplicative depth. \\[2pt]

% $N$, $Q$, $\Delta$ &
% Polynomial degree; ciphertext modulus; CKKS scaling factor. \\[2pt]

$D$, $d_i$, $q_i$, $Q_i$, $u_i$ &
Integer; its residue; $i$-th RNS prime; CRT factor $Q_i=Q/q_i$; modular inverse $u_i=Q_i^{-1}\bmod q_i$. \\[2pt]

$e$, $E$ &
Single-bit local error or noise; amplified error after computation. \\[2pt]

$\mathsf{Flag}_{\text{in}}$, $\mathsf{Flag}_{\text{out}}$ &
Input/output integrity flags for checksum-based fault-tolerance algorithms.\\
%FHE-Poly-Check fault detection. \\

\bottomrule
\end{tabular}
\caption{Main notation used in this paper.}
\label{tab:notation}
\vspace{-1cm}
\end{table}

\section{Exploration of SDC in CKKS 
Evaluation}
\label{sec_Vulnerability Evaluation}

Sec.~\ref{fault_model} presents the fault model adopted in our evaluation, and Sec.~\ref{sec:Vulnerability Evaluation for FHE OPs} and~\ref{sec: Detailed Evaluation in keyswitch} then analyze the resulting hardware-induced SDC behaviors in CKKS via fault-injection experiments.

\subsection{Fault Model}
\label{fault_model}

The primary goal of this work is to study how faults occurring during ciphertext computation propagate and affect the final decrypted message.
Accordingly, our evaluation adopts the following principles.
First, because we target silent data corruptions (SDCs) in ciphertext-based services, we inject faults only during the ciphertext computation phase; the resulting ciphertext is then decrypted fault-free, and the deviation from the correct message is recorded.
Second, since our objective is to characterize fault-propagation behavior, we assume that at most one transient fault occurs per execution, and we measure how this single perturbation influences the decrypted result.

To emulate such faults, we adopt the widely used \emph{random bit-flip} abstraction~\cite{zhang2023read,sangchoolie2017onebit,he2020fidelity,hsiao2023ros_sdc,papadimitriou2021vulnerabilitystack}.
Specifically, we inject single-bit flips into the instruction stream of CKKS ciphertext processing.
This single-bit transient fault model is commonly used in reliability studies and serves as a minimal-perturbation baseline for analyzing fault-propagation behavior.

\begin{figure*}[htbp]
    \centering
    \includegraphics[width=0.99\linewidth]{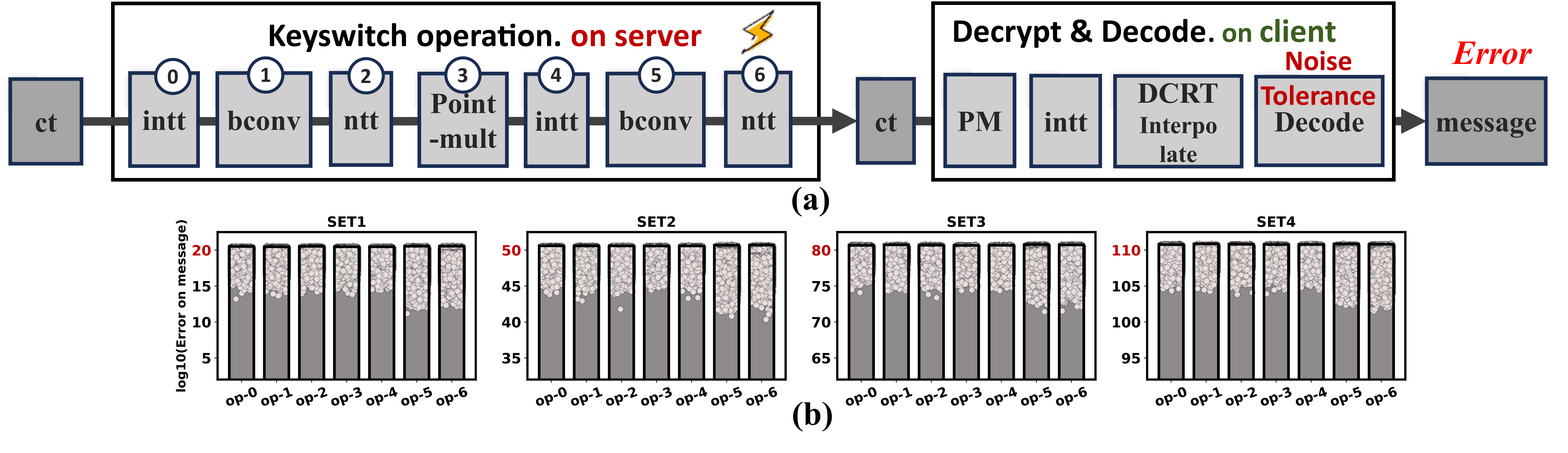}
    \vspace{-0.5cm}
    \caption{(a) Seven polynomial operation steps in Keyswitch: op-0: intt, op-1: bconv, op-2: ntt, op-3: point-mult, op-4: intt, op-5: bconv, op-6: ntt. The PM in Decrypt denotes point-mult. (b) Error evaluation on the decrypted message after injecting faults on each step.}
    \label{fig:fhe_step}
\end{figure*}

\subsection{Vulnerability Evaluation on Single Basic Ciphertext Operations}
\label{sec:Vulnerability Evaluation for FHE OPs}

% Figure~\ref{fig:fhe_operator}(a)–(b) presents the vulnerability analysis of five basic CKKS homomorphic operations under single-bit fault injection. 
While Fig.~\ref{fig:fhe_app}(b) has already shown that CKKS-based encrypted AI inference suffers severe accuracy degradation under transient faults, this section further analyzes the SDC vulnerability of the fundamental CKKS operators themselves.

\paragraph{(1) Operator set.}
CKKS ciphertext computation follows a static dataflow. 
Each high-level ciphertext function consists of several basic homomorphic operators. 
We evaluate the five most fundamental operators, which together form the minimal operation set of CKKS:
{\small
\begin{itemize}
    \item \texttt{ct-pt mult}: ciphertext–plaintext multiplication,
    \item \texttt{ct-ct mult}: ciphertext–ciphertext multiplication,
    \item \texttt{ct-pt add}: ciphertext–plaintext addition,
    \item \texttt{ct-ct add}: ciphertext–ciphertext addition,
    \item \texttt{ct rot}: ciphertext rotation.
\end{itemize}}
\paragraph{(2) Parameter configurations.}
Four CKKS parameter settings are used (SET1–SET4, see Fig.~\ref{fig:fhe_operator}(a)), corresponding to multiplicative depths $L=\{2,4,6,8\}$.  
For all settings, the ciphertext dimension is $N=32768$. The ciphertext modulus $Q$ are configured to 180, 280, 380, 480 bits, respectively, 
while the scaling factor $\Delta$ is set to 50~bits.  
All ciphertexts are generated using full-slot packing, and the input message and the resulting message after decryption are full-packed vectors.
\paragraph{(3) Fault injection and Evaluation.}
Each CKKS operator is implemented using the widely used library, OpenFHE ~\cite{OpenFHE} and compiled into an independent executable. The experiments are run on an Intel(R) Xeon(R) Platinum 8358P CPU.
Random single-bit transient faults are injected into ciphertext computations using the Intel Pin dynamic instrumentation framework~\cite{intel_pin_tool}.  
Each program is executed 10,000 times, with exactly one random bit flipped per run. %As widely adopted in prior fault-injection studies, single-bit transient faults serve as the minimal perturbation model for analyzing fault propagation behavior.
\paragraph{(4) Experimental results and analysis.}
The results show that all CKKS operators exhibit high vulnerability to transient faults. Across all operators and parameter sets, approximately 22\% of injected single-bit faults lead to SDCs.
Compared with crash failures, SDCs pose a more severe system-level threat because they produce erroneous ciphertext outputs without any observable failure symptoms.
To characterize the severity of these SDCs, we further examine the magnitude of output deviations.
Since CKKS decryption returns a message vector, we measure, for each SDC case, the numerical deviation of every message slot from the fault-free result.
These results are visualized in Fig.~\ref{fig:fhe_operator}(b), where each subfigure corresponds to one parameter configuration (SET1–SET4), and each bar represents one CKKS operator.
Each point in a bar denotes the deviation of an individual message slot relative to the correct output.
% Across all operators and parameter sets, about 25\% of the injections lead to SDCs.
% 相比于crash,SDC对系统有更严重的威胁.
% 我们进一步分析SDC出现时,对应的误差大小. 由于CKKS解密的输出是消息向量,我们统计解密结果message的每个slot的误差有多大.
% 结果展示在Figure~\ref{fig:fhe_operator}(b)中, where each subfigure corresponds to one parameter configuration (SET1–SET4), 
% and each bar represents one CKKS operator. Each point within a bar denotes the distance for each slot of the message compared with fault-free computation.

As shown in Fig.~\ref{fig:fhe_operator}(b), \textbf{first}, once an SDC occurs, \emph{every} slot in the decrypted message typically incurs a large deviation.
This indicates that hardware-induced SDCs in ciphertext computation almost always translate into \emph{substantial corruption} at the application level.
\textbf{Second}, the magnitude of slot-level errors is strongly influenced by the ciphertext modulus~$Q$.
Across the four CKKS parameter sets, the maximum deviation grows monotonically with~$Q$: when $Q$ increases from 180 to 480 bits (SET1–SET4), the upper bound of slot errors expands from approximately $10^{35}$ to $10^{125}$—an increase of roughly $10^{30}$ per additional 100 bits of modulus.
% This trend reflects the fact that a larger modulus provides a larger dynamic range for error amplification through the (I)NTT and DCRT interpolation steps.
Within each parameter set, multiplication-related operators (\texttt{ct-pt mult} and \texttt{ct-ct mult}) consistently produce smaller deviations than addition and rotation.
This is because multiplication triggers a \texttt{rescale} operation, which consumes part of the ciphertext modulus and reduces the remaining modulus available for subsequent error amplification.

\subsection{Fault Injection in Polynomial Operation Steps in Basic FHE Operators}
\label{sec: Detailed Evaluation in keyswitch}

% Further, each ciphertext operator itself consists of a sequence of polynomial operations.  
% Analyzing faults at this granularity helps reveal how errors propagate across different computational layers and provides guidance for designing fault-tolerant algorithms at the polynomial level.
% Therefore, in this section, we focus on the polynomial-level operators. We evaluate different polynomial operator steps in \texttt{KeySwitch}, one of the most critical and computation-intensive primitives in CKKS.  
% \texttt{KeySwitch} is the main computation step in both \texttt{HMult} and \texttt{HRot}, and it accounts for more than 90\% of the runtime in practical ciphertext applications.
% As illustrated in Fig.~\ref{fig:fhe_step}(a), the \texttt{KeySwitch} procedure mainly consists of seven polynomial operation steps connected in a static pipeline.

Furthermore, since each ciphertext-level operator is composed of a sequence of polynomial-level primitives, we further examine the SDC vulnerability of these internal steps.
Specifically, we focus on \texttt{KeySwitch}, one of the most critical and computation-intensive primitives in CKKS.
It serves as the core computation step in both \texttt{HMult} and \texttt{HRot}, and accounts for more than 90\% ~\cite{CKKS} of the overall runtime.
As illustrated in Fig.~\ref{fig:fhe_step}(a), the \texttt{KeySwitch} procedure consists of seven polynomial operations connected in a fixed pipeline.

\paragraph{(1)Experimental setup.}
% The experiment focuses on the seven polynomial level operations inside the \texttt{Keyswitch} procedure, which is internally triggered during a ct-ct mult.
The experiment focuses on the seven polynomial-level operations of the \texttt{KeySwitch} procedure as invoked during a ct-ct mult.
The crypto parameters are under the same four configurations (SET1–SET4) described in Sec.~\ref{sec:Vulnerability Evaluation for FHE OPs}.
The fault-injection methodology is identical—each execution introduces exactly one random single-bit fault using Intel Pin~\cite{intel_pin_tool}.
Each program is executed 10,000 times.

\paragraph{(2) Results and analysis.}

Across all seven polynomial steps, the observed SDC rates are consistently over 20\%, indicating that every polynomial step exhibits high vulnerability.
To quantify the severity of these SDCs, we record the per-slot deviation of the decrypted message vector for each SDC case.  
The results for all four parameter settings and all seven steps are shown in Fig.~\ref{fig:fhe_step}(b), where each bar corresponds to one polynomial step.  
As shown in Fig.~\ref{fig:fhe_step}(b), transient faults occurring at different steps can lead to \emph{large deviations across all slots} of the decrypted message.
Furthermore, comparing results across the four parameter settings reveals that the error magnitude again correlates strongly with the ciphertext modulus~$Q$, with larger~$Q$ enabling greater amplification of hardware-induced perturbations.

\section{Error Tolerance Algorithm Evaluation}

This section evaluates algorithm-level mechanisms for tolerating transient faults in CKKS ciphertext computation. 
We study two representative protection strategies:
(1) redundant-based execution, and 
(2) checksum-based verification applied to polynomial-domain operators. 

% Section~\ref{sec:polycheck-design} introduces the approaches, and Section~\ref{sec:polycheck-eval} evaluates the fault coverage and performance overhead of both methods under single-bit fault injection scenarios in FHE computations.

% 上述分析说明了FHE容错算法的必要性。我们在本章展示了容错机制的评估。
% 由于FHE计算特殊，且FHE容错算法的研究较少，我们主要展示了2种算法：一种是重计算，一种是我们设计的FHE-poly-check。
% 我们在第一节介绍了FHE-poly-check，在第二节评估了他们对FHE计算中单bit故障的覆盖率和性能开销。

% Given the high vulnerability of FHE ciphertext computation to hardware faults, efficient fault-detection mechanisms are urgently needed.  
% While memory faults can be effectively mitigated by ECC, we propose a lightweight detection framework, termed \textbf{DF-FHE}, to efficiently identify and recover from computation faults within ciphertext operations.

% \subsection{Design of FHE-Poly-check}
% FHE计算是由多项式运算组成的静态计算流，因此we design a \textbf{checksum-based verification algorithm} for polynomial computations in FHE. \textbf{Once an error is detected within any polynomial computation, that specific polynomial operation is re-executed to achieve error correction.}
% 值得一提的是，we integrate a \textbf{lazy reduction} technique to minimize the modular reduction cost in the checking process.

\subsection{Error Tolerance Algorithm}
\label{sec:polycheck-design}

As discussed in Sec.~\ref{sec:CKKS-prelim}, CKKS evaluation follows a static, data-oblivious pipeline composed of polynomial operations such as (I)NTT, pointwise multiplication, basis conversion (BConv), and DCRT interpolation.  
Based on this computation pipeline, we evaluate two baseline fault-mitigation algorithms: 1) Redundant-Based, and 2) Checksum-Based.
For the checksum-based verification, we did not design new ABFT schemes; we integrated and adapted existing methods from matrix multiplication, FFT/NTT, and RNS arithmetic into the FHE computation flow.

\paragraph{1) Redundant-Based.}
For each polynomial operator, dual-modular redundancy (DMR) executes the operator twice and compares the results.  
A mismatch indicates an error, and the operator is re-executed for correction.  
% This provides full coverage at the cost of approximately one additional execution per protected operator.

\paragraph{2) Checksum-Based.}
Checksum-based ABFT techniques verify input--output consistency through lightweight linear invariants.  
We apply known ABFT schemes to CKKS polynomial operators.
For example, for NTT/INTT, we adopt the technique of~\cite{abdelmonem2025efficient}:  since $\mathrm{NTT}(a)=W a$ with twiddle-factor matrix $W$, a vector $F$ can be chosen such that $F W = \{1\}^{N}$, enabling verification via
\begin{equation}
\scriptsize
\mathsf{Flag}_{\text{in}}=\sum_{i=0}^{N-1} a_i \bmod q,
\qquad
\mathsf{Flag}_{\text{out}}=\sum_{i=0}^{N-1} A_i F_i \bmod q.
\end{equation}
detecting any single-bit transient fault within the NTT computation.
Similar checksum structures are extended to other polynomial operators to fit into the FHE pipeline.
These techniques reload input and output data, enabling detection of a subset of load/store-related transient faults as well.

% BConv map an RNS vector $\{x[m]_{q_i}\}$ to another basis $\{x[m]_{p_j}\}$. For BConv, we extend existing RNS ABFT checks to multi-limb input/output:
% \begin{equation}
% \scriptsize
% \mathsf{Flag}_{\text{in}}
% =
% \left(
% \sum_{i=0}^{l-1}
% \hat{q}_i 
% \cdot 
% \Bigl(
% \sum_{m=0}^{N-1} x[m]_{q_i}
% \Bigr)
% \cdot 
% \hat{q}_i^{-1} \bmod q_i
% \right),
% \quad
% \mathsf{Flag}_{\text{out}}
% =
% \left(
% \sum_{j=0}^{k-1}
% \sum_{m=0}^{N-1} x[m]_{p_j}
% \right).
% \end{equation}
% A mismatch indicates an error during BConv.

\subsection{Evaluation Setup}

We evaluate four representative ciphertext-level workloads: vector–vector multiplication (VV), matrix–vector multiplication (MV), ciphertext rotation (Rot), and a regression task on the California Housing dataset~\cite{nugent_california} (House).
The ciphertext parameter settings are identical to those of SET 3 in Sec.~\ref{sec_Vulnerability Evaluation}.
We compare three execution modes: None (no protection), redundant-based, and checksum-based.
Each program is executed for 50,000 runs, and in each run a transient single-bit fault is injected using \texttt{PinFI}~\cite{intel_pin_tool}.
All experiments are performed on an Intel(R) Xeon(R) Platinum 8358P CPU.
 
% We evaluate: vector--vector multiplication (VV) in ciphertext,  matrix--vector multiplication (MV) in ciphertext, ciphertext rotation (Rot), and a regression task (California Housing~\cite{nugent_california}, House) in ciphertext.
% We compare three execution modes: None (no protection), redundancy-based, and checksum-based.
% Each program is executed for 10,000 runs. 每次执行使用\texttt{PinFI}~\cite{intel_pin_tool}注入Transient single-bit faults. All experiments are performed on an Intel(R) Xeon(R) Platinum 8358P CPU.   

\begin{figure}[htbp]
    \centering
    \includegraphics[width=0.8\linewidth]{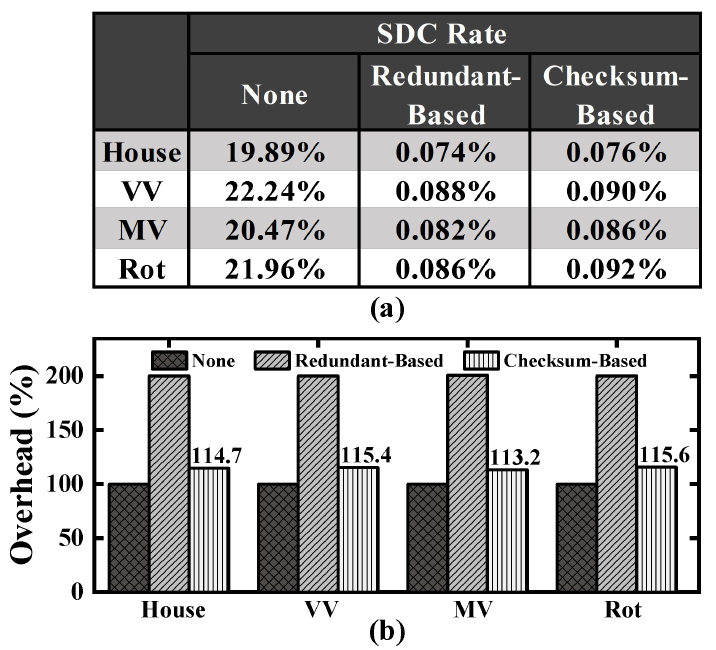}
    %\includegraphics[width=0.8\linewidth]{fig/fig_ft_1119.png}
     % \includesvg[inkscapelatex=false,width=0.8\linewidth]{fig/fig6.svg}

    \caption{(a) SDC rate evaluation. (b) The overhead of fault-tolerant ciphertext computation.}

    \label{fig:runtime_overhead_final}
\end{figure}

\subsection{Results and Analysis}

Fig.~\ref{fig:runtime_overhead_final}(a) reports the end-to-end SDC rate under single-bit transient faults.  
Across all workloads, unprotected CKKS evaluation exhibits a SDC rates of about $19.89\%-22.24\%$.
The redundant-based approach reduces the SDC rate to below 0.1\%, while the checksum-based method achieves a comparable level of protection.

Fig.~\ref{fig:runtime_overhead_final}(b) presents the runtime overhead. 
We normalize the fault-free execution time to $1$ and report the total latency of the two fault-tolerant schemes relative to this baseline. 
As shown in the figure, redundant-based execution introduces approximately a $1\times$ slowdown, while checksum-based detection incurs an overhead of about $13$--$16\%$.

% Figure~\ref{fig:runtime_overhead_final} (b) presents the runtime overhead. 我们记录无防护的原始执行时间为单位1,然后展示了2种带容错算法方案的总延迟.如图所示,
% Redundant-based execution introduces approximately a $1\times$ slowdown. Checksum-based detection incurs an overhead of approximately $13$--$16\%$.

\subsection{Summary}

% 从实验结果可以看到classcial DMR 可以有效降低SDC rate，但是恶化本来就很重FHE的计算开销。相比之下，checksum方法更加轻量化一些，只是引入了约$15\%$的开销，能够达到与classcial DMR接近的保护效果。

The experimental results show that classical DMR effectively reduces the SDC rate, but it significantly increases the already substantial computational cost of FHE. In contrast, the checksum-based method is more lightweight, introducing only about a $15\%$ overhead while achieving a protection level comparable to classical DMR. Nevertheless, given the extremely high computational cost of FHE, even this level of overhead is nontrivial, and developing more lightweight SDC-tolerance strategies remains an important direction for future work.

\section{Conclusion and Future Work}

% In this paper, we investigate an interesting and important problem: SDCs in FHE ciphertext computation. 
% We perform large-scale fault-injection experiments and show the high vulnerability of FHE ciphertext computation. 
% We further analyze the computation patterns of FHE and reveal how small errors are amplified in both slot-level diffusion and value magnitude. 这使得我们对FHE对SDC的脆弱性的算法曾层机制有了更深入认知.
% Finally, we evaluate two types of fault-tolerant algorithms.
% It should be noted that this remains an important problem requiring further study. 
In this paper, we investigate an interesting and important problem: SDCs in FHE ciphertext computation.
We perform large-scale fault-injection experiments and demonstrate the high vulnerability of FHE evaluation to transient hardware faults.
We further analyze the computation patterns of FHE and reveal how small perturbations are structurally amplified through slot-level diffusion and full-modulus error growth, providing deeper algorithmic insight into why FHE is inherently fragile to SDCs.
Finally, we evaluate two types of fault-tolerant algorithms.
This study contributes to advancing the practical deployment of FHE and improving the robustness of the privacy-preserving computing ecosystem. We hope that our findings draw broader attention to the reliability of privacy-preserving computation when deployed on real hardware platforms.
Looking forward, we plan to further explore lightweight and hardware-aware fault-tolerant FHE computation. Our results indicate that every step of CKKS evaluation requires reliable protection. We will investigate how to jointly optimize reliability and performance on real hardware and design tailored, low-cost fault-tolerance mechanisms for the FHE pipeline. 

\newpage

%\bibliographystyle{ACM-Reference-Format}
%\bibliography{reference}

% \appendix









\end{document}